\documentclass[published,notoc]{JHEP3} 

\JHEP{08(2001)057}

\usepackage{epsfig,multicol}

\usepackage{latexsym}

\newcommand\fverb{\setbox\pippobox=\hbox\bgroup\verb}
\newcommand\fverbdo{\egroup\medskip\noindent%
			\fbox{\unhbox\pippobox} }
\newcommand\fverbit{\egroup\item[\fbox{\unhbox\pippobox}]}
\newbox\pippobox

\def\plushc{ + h.c.}

\def\sgn{{\rm sgn}}
\newcommand{\nn}{\nonumber}

\def\be{\begin{equation}}
\def\ee{\end{equation}}
\def\bea{\begin{eqnarray}}
\def\eea{\end{eqnarray}}
\def\Ref#1{(\ref{#1})}
\def\cropen#1{\crcr\noalign{\vskip #1}}
\def\crr{\cropen{1\jot }}
\newcommand{\rc}{r}
\newcommand{\keff}{\bar{\kappa}}
\newcommand{\barR}{\bar{R}}
\newcommand{\bare}{\bar{e}}
\newcommand{\barg}{\bar{g}}
\newcommand{\barh}{\bar{h}}
\newcommand{\barT}{\bar{T}}
\newcommand{\I}{{\rm i}}
\newcommand{\hfive}{{\hat 5}}

\title{Supersymmetric radion in the \\
Randall-Sundrum scenario}

\author{Jonathan Bagger\\
Department of Physics and Astronomy, The Johns Hopkins University,\\
3400 North Charles Street, Baltimore, MD 21218\\
E-mail: \email{bagger@jhu.edu}}

\author{Dennis Nemeschansky\\
CIT-USC Center for Theoretical Physics and Dept. of Physics and Astronomy,\\
University of Southern California, Los Angeles, CA 90089 \\E-mail: \email{dennisn@physics.usc.edu}}

\author{Ren-Jie Zhang\\
School of Natural Sciences, Institute for Advanced Study, \\
Olden Lane, Princeton, NJ 08540\\
E-mail: \email{rjzhang@ias.edu}}

\received{July 27, 2001} 		
\accepted{August 29, 2001}		

\abstract{
We derive the effective action for the radion supermultiplet
in the\break supersymmetric Randall-Sundrum model with two opposite
tension branes.}

\begin{document} 


\section{Introduction}

The two-brane Randall-Sundrum scenario \cite{RS} has attracted
considerable attention because it generates the electroweak
hierarchy as a consequence of spacetime geometry.  The simplest
version contains two opposite tension 3-branes placed at the
boundaries of a slice through 4+1 dimensional anti de
Sitter space.  The hierarchy is determined by the brane tensions
and by the distance between the branes.  The distance is set
by the expectation value of a modulus field, called the radion.
Supersymmetry provides a natural mechanism for stabilizing
the radion against radiative corrections.

During the past year, supersymmetric versions of the two-brane
Randall-Sundrum scenario were constructed \cite{ABN,others}.
Furthermore, the corresponding low-energy effective supergravity
action was derived --- for the case of fixed radion.
In this paper we will find the effective field theory for the
field radion itself.  We will derive the $N=1$ matter-coupled
supergravity theory that describes the low energy dynamics of
the radion.

The plan of this paper is as follows.  In section 2 we review
the five dimensional bulk-plus-brane supersymmetric action.  In
section 3 we derive the bosonic part of the low energy effective
action.  We will see that the Kaluza-Klein reduction requires
a careful treatment of tadpoles associated with the massive fields.
In section 4 we construct the fermionic part of the effective
action, and in section 5 we derive the supersymmetry transformations
for the zero mode fields.  We will find that the
lagrangian and transformation laws
correspond to the usual $N=1$ matter-coupled supergravity, with
a K\"ahler potential of a particular form.  We conclude in
section 6 with a simple argument that produces the same K\"ahler
potential.

\section{The five dimensional supersymmetric action}

Our starting point is the supersymmetric Randall-Sundrum model
presented in \cite{ABN}.  The total action contains bulk and brane
pieces,
\be
S = S_{\rm bulk} + S_{\rm brane} .
\label{totalS}
\ee
The bulk action is that of five dimensional supergravity
\cite{AdS}, while $S_{\rm brane}$ arises from the presence of
two opposite-tension branes.\footnote{Our final results are
independent of whether we start with the five dimensional
action presented in Ref.~\cite{ABN} or \cite{others}.}

The bulk action is given by
\bea
S_{\rm bulk} & = & \Lambda  \int d^5x
 e \bigg[  - {1 \over 2 \kappa^2} R
 + \I \epsilon^{MNOPQ}
\overline\Psi_{M}\Sigma_{NO} D_P \Psi_Q
 - {1 \over 4} F_{MN} F^{MN} \,-\nn \\
& & -  3  \Lambda  \bar\Psi_M\Sigma^{MN}\Psi_N
 + 6  {\Lambda^2\over\kappa^2}
 - \I \kappa {\sqrt{3\over2}}{1\over2}
F_{MN}\bar\Psi^M\Psi^N\, - \nn\\
& &- \kappa {1\over 6\sqrt{6}}
\epsilon^{MNOPQ}F_{MN}F_{OP}B_Q
 + \I \kappa \sqrt{3\over2}{1\over 4}
\epsilon^{MNOPQ}F_{MN}
\bar\Psi_O \Gamma_P\Psi_Q \, -\nn\\
& & - \kappa  \Lambda {\sqrt{3\over 2}}
\epsilon^{MNOPQ}
\bar\Psi_M\Sigma_{OP}\Psi_N B_Q
 +  \hbox{ four-Fermi terms } \bigg] .
\label{bulkS}
\eea
In this expression, the parameter $\kappa$ is related to the
effective four dimensional Planck constant, $\kappa^2 = \keff^2
(1 - e^{-2 \pi \rc\Lambda})$, and the $\epsilon$ tensor is defined
to have tangent-space indices,\footnote{We adopt the convention
that capital letters run over the set $\{0,1,2,3,5\}$ and lower-case
letters run from 0 to 3.  Tangent space indices are taken from the
beginning of the alphabet; coordinate indices are from the middle.
We follow the conventions of \cite{ABN,WB}.}
with $\epsilon^{01235} = 1$.  The action contains the physical fields
of the supergravity multiplet in five dimensions:  the f\"unfbein
$e_M{}^A$, the gravitino $\Psi_M$, and the graviphoton $B_M$.
The coordinate $x^5 = z = \rc \phi$ parameterizes the orbifold $S^1/Z_2$,
where the circle $S^1$ has radius $\rc$ and the orbifold identification
is $\phi \leftrightarrow -\phi $. We choose to work on the orbifold
covering space, so we take $-\pi < \phi \le \pi$.  The orbifold breaks
$N=1$ supersymmetry in five dimensions to $N=1$ in four.  

The brane action is intrinsically four dimensional, so we write the
five dimensional spinors in four dimensional notation, where
\be
\Psi_M  = \pmatrix{ \psi^1_{M \alpha} \crr
\bar\psi^{\dot\alpha}_{2 M}}
\ee
and
\be
\Gamma^a  = \pmatrix{ 0 &
\sigma^a_{\alpha\dot\alpha} \crr
\bar\sigma^{a\dot\alpha\alpha} & 0} \qquad \qquad
\Gamma^\hfive = \pmatrix{ -\I & 0 \crr 0 & \I} .
\ee
The symbol $\hfive$ denotes the fifth tangent space index;
the fields $\psi^i_M$ are two-component Weyl spinors.  We define
$\psi^\pm_M = {1\over \sqrt 2}(\psi^1_M \pm \psi^2_M)$, and likewise
for the supersymmetry transformation parameter $\eta^\pm$.

In this language, the brane action is simply
\be
S_{\rm brane}  = {\Lambda \over \kappa^2 }
  \int d^5x  \hat{e}
 (-3 \Lambda
 + 2 \kappa^2  \psi^+_m \sigma^{mn} \psi^+_n)
 \left[ \delta(z) - \delta(z-\pi\rc)  \right]
\plushc ,
\label{braneS}
\ee
where $\hat{e} = \det e_m{}^a$ and the $e_m{}^a$ are the components
of the f\"unfbein, restricted to the appropriate brane.

The full bulk plus brane action
is invariant under the following transformations,
\bea
\delta e_M{}^a &=& \I\kappa  (\eta^+\sigma^a\bar\psi^+_{M}
  +  \eta^-\sigma^a\bar\psi^-_{M})
\plushc \nn\\[2mm]
\delta e_M{}^\hfive &=& \kappa  (\eta^+\psi^-_{M}
  -  \eta^-\psi^+_{M})
\plushc \nn\\
\delta B_M &=& -\I\sqrt{3\over 2} 
(\eta^+\psi^-_{M}
  -  \eta^-\psi^+_{M})  \plushc  \nn\\
\delta \psi^\pm_m &=& {2\over\kappa} D_m\eta^\pm
 \mp {\I\over\kappa}  \omega_{ma\hfive}  \sigma^a
\bar\eta^\mp  \pm {\I} {\Lambda\over\kappa}
 e_m{}^a  \sigma_a \bar\eta^\pm
 +  e_{m\hfive} {\Lambda\over\kappa} \eta^\mp \, -\nn\\
&& - \I \sqrt{6} \Lambda  B_m  \eta^\mp
 - \sqrt{2\over 3} \bigg(\mp  e_a{}^N F_{Nm} 
\sigma^a\bar\eta^\mp - \I  e_\hfive{}^N  F_{Nm}  \eta^\pm \, -\nn\\
&& - {1\over 4}  \epsilon_{ABCde} 
e_m{}^A e^{BN} e^{CO}  F_{NO}  \sigma^{de} \eta^\pm
 \pm {\I\over 4}  \epsilon_{abcd} 
e_m{}^a e^{bN} e^{cO}  F_{NO}  \sigma^{d} \bar\eta^\mp \bigg)
\nn\\[2mm]
\delta \psi_5^+ &=& {2\over\kappa} D_5\eta^+
 - {\I\over\kappa}  \omega_{5a\hfive}
 \sigma^a \bar\eta^-
 + e_{5\hfive} {\Lambda\over\kappa}  \eta^-
 + \I e_5{}^a {\Lambda\over\kappa} 
\sigma_a \bar\eta^+ \, -\nn\\
&& - \I\sqrt{6} \Lambda  B_5  \eta^-
 - \sqrt{2\over 3} \bigg(  e_a{}^n F_{5n} 
\sigma^a\bar\eta^- + \I  e_\hfive{}^n  F_{5n}  \eta^+\, - \nn\\
&& - {1\over 4}  \epsilon_{ABCde} 
e_5{}^A e^{BN} e^{CO}  F_{NO}  \sigma^{de} \eta^+
 - {\I\over 4}  \epsilon_{abcd} 
e_5{}^a e^{bN} e^{cO}  F_{NO}  \sigma^{d} \bar\eta^- \bigg)\nn\\[2mm]
\delta \psi_5^- &=& {2\over\kappa} D_5\eta^-
 + {\I\over\kappa}  \omega_{5a\hfive}
 \sigma^a \bar\eta^-
 + e_{5\hfive} {\Lambda\over\kappa}  \eta^+
 - \I e_5{}^a {\Lambda\over\kappa} 
\sigma_a \bar\eta^- \, -\nn\\
&& - \I\sqrt{6} \Lambda  B_5  \eta^+
 + \sqrt{2\over 3} \bigg(  e_a{}^n F_{5n} 
\sigma^a\bar\eta^+ - \I  e_\hfive{}^n  F_{5n}
  \eta^- \, +\nn\\
&& + {1\over 4}  \epsilon_{ABCde} 
e_5{}^A e^{BN} e^{CO}  F_{NO}  \sigma^{de} \eta^-
 - {\I\over 4}  \epsilon_{abcd} 
e_5{}^a e^{bN} e^{cO}  F_{NO}  \sigma^{d} \bar\eta^+
\bigg)\, - \nn\\
&& - {4\over \kappa}
  [ \delta(z)  -  \delta(z-\pi\rc) ] 
\eta^+  ,
\label{susytrans}
\eea
provided the fields and transformation parameters obey the
appropriate jump conditions at the locations of the branes.
In these
expressions, all covariant derivatives contain the spin connection
$\omega_{Mab}$.  
Here and hereafter, we ignore all three- and four-Fermi terms.

The fields and transformation parameters are assigned definite
$Z_2$ parities under the orbifold symmetry $\phi \rightarrow -\phi$.
We choose
$$
e_m{}^a, \quad e_{5\hfive}, \quad B_5, \quad \psi^+_{m}, \quad
\psi^-_{5}, \quad \eta^+
$$
to have even parity, and
$$
e_5{}^a, \quad e_m{}^\hfive, \quad B_m, \quad \psi^-_{m}, \quad
\psi^+_{5}, \quad \eta^-
$$
to have odd.  These parity assignments
are consistent with the action and the supersymmetry transformations.

\section{The bosonic effective action}

In this paper we derive the effective field theory for the
zero modes associated with the five dimensional supergravity fields.
We consider the four dimensional $N=1$ supergravity multiplet,
as well as the four dimensional chiral multiplet that contains
the radion.  In this section we find the bosonic part of the
low energy effective action.

We take the vacuum to be the original
Randall-Sundrum solution \cite{RS},
\be
ds^2 = e^{- 2 \sigma}  \eta_{mn}  dx^m dx^n + dz^2 .
\label{RSmetric}
\ee
This solves the five dimensional Einstein equations
\be
R_{MN} - {1\over2}  g_{MN} R = -6   g_{MN} \Lambda^2 +
 6  g_{mn} \delta^m_M
\delta^n_N   \Lambda
\left( {\hat e \over e}\right)
[ \delta(z)  -  \delta(z-\pi\rc) ]
\label{fiveeinstein}
\ee
when $\eta_{mn}$ is the flat Minkowski metric and $\sigma =
\Lambda |z|$.

The low energy effective action describes the  light
fields that fluctuate off this background.  The effective theory
of the gravitational field was derived in Ref.~\cite{RS}.  The
zero mode metric is
\be
ds^2 = e^{- 2\sigma}  \barg_{mn}  dx^m dx^n + dz^2 ,
\label{gzero}
\ee
where $\barg_{mn}$ is the effective four dimensional metric, a function
of $x^0,...,x^3,$ but not $x^5$.  The effective action follows from
substituting \Ref{gzero} into \Ref{totalS} and integrating over $x^5$.
One finds
\be
S_{\rm eff}  = -  {1 \over 2 \keff^2}\int d^4x
 \bare \barR ,
\label{Seff}
\ee
where $\barR$ is the four dimensional Ricci scalar constructed
from the metric $\barg$.  This is nothing but the Einstein action
in four dimensions, with an effective four dimensional squared Planck
mass $\keff^{-2} = \kappa^{-2}(1 - e^{-2 \pi \rc\Lambda})$.

This procedure is consistent because the five
dimensional Einstein equations \Ref{fiveeinstein} are
satisfied when
\be
\barR_{mn} = 0 .
\ee
Indeed, the ansatz \Ref{gzero} gives a solution,
point-by-point in $x^5$, precisely when $\barg_{mn}$
satisfies the four dimensional Einstein equations,
derived from the effective
action \Ref{Seff}.  This implies that the Kaluza-Klein reduction
is consistent; all higher Kaluza-Klein modes can be set
to zero.

In what follows we generalize this reduction to include the
radion field, $A$.  We construct a four dimensional effective
action that is correct to all orders for constant radion $A$,
and correct up to two derivatives when $A$ depends on the four
dimensional coordinates $x^0,...,x^3$.  The resulting effective action
describes the leading low energy dynamics of the radion
field.

We start by writing a five dimensional metric that satisfies
the Einstein equations for constant $A$.  Such a metric
is given by
\be
ds^2 = e^{- 2 F(A,\sigma)}  \barg_{mn}  dx^m dx^n
 + \left|{\partial F(A,\sigma)\over \partial \sigma}
\right|^2  dz^2 .
\label{gA}
\ee
For constant $A$, \Ref{gA} can be obtained from \Ref{gzero} by
a coordinate transformation, so it automatically satisfies
the five dimensional equations of motion.

When the field $A$ depends on the four dimensional coordinates
$x^0,...,x^3$, the metric \Ref{gA} is not a coordinate transformation
of \Ref{gzero}, and the reduction is more complicated.  The problem is
that $A$ mixes with the graviton $\barg_{mn}$ and
with its Kaluza-Klein excitations $\barh_{mn}$.  Moreover, the
amount of mixing typically depends on the coordinate $x^5$.  This
makes it subtle to extract the four dimensional effective theory.

These problems are eliminated by choosing the
following ansatz for the five dimensional
metric\ \footnote{This metric coincides,
to linear order, with the metric proposed in \cite{RG}.}
\be
ds^2 = e^{- 2 \sigma}  \barg_{mn}  
(1 + A  e^{2 \sigma})  dx^m dx^n
 + {1\over (1 + A  e^{2 \sigma})^2}  dz^2 .
\label{Ametric}
\ee
This metric is of the form \Ref{gA}, where
\be
F(A,\sigma) = \sigma  - {1\over2} \log  (1 +
A e^{2 \sigma}) .
\ee
With this ansatz, the gravitational part of the five dimensional
action \Ref{totalS} is just
\be
S  =  \Lambda  \int d^5x
 \bare \bigg[ - {1 \over 2 \kappa^2} e^{-2 \sigma} \barR
 - {3 \over 4 \kappa^2} {e^{2 \sigma}\over (1 + A  e^{2 \sigma})^2}
 \barg^{mn}  \partial_m A   \partial_n A \bigg] .
\label{Azero}
\ee
Equation \Ref{Azero} is in the ``Einstein frame" at each point in $x^5$.  
This guarantees that the radion never mixes with the graviton, to any
order in the fields.

The radion effective action follows from integrating \Ref{Azero}
over $x^5$.  This gives
\bea
S_{\rm eff} &=& \int d^4x  \bare \bigg[ - {1 \over 2 \keff^2}
\barR - {3 \over 4 \keff^2}  {e^{2\pi\rc\Lambda} \over (1 + A)(1 + A
 e^{2 \pi\rc\Lambda})} \barg^{mn}  \partial_m A   \partial_n A \bigg]
\nn\\[2mm]
&=& \int d^4x  \bare \bigg[ - {1 \over 2 \keff^2}
\barR - {1 \over 8 \keff^2 } \sinh^{-2}\bigg({ B\over \sqrt{6}} \bigg)\
\barg^{mn}  \partial_m B   \partial_n B \bigg] ,
\label{SeffR}
\eea
where we write $A$ in terms of the $x^5$-invariant proper length, $B$,
\be
\sqrt{{2\over3}}  B = 2 \Lambda 
\int_0^{\pi\rc} e_{5\hfive} dz = 2\pi\rc\Lambda
 +  \log 
\bigg[ {1+A\over 1 + A  e^{2\pi\rc\Lambda}} \bigg] .
\ee
The field $B$ describes the physical distance between
the opposite-tension branes.

Let us now examine the consistency of this reduction.  The four dimensional
equations of motion are easily derived from \Ref{SeffR}.  We find
\be
\barR_{mn}  =\
- {1 \over 4} \sinh^{-2}\bigg({ B\over \sqrt{6}} \bigg)
 \partial_m
B \partial_n B
\label{fourd1}
\ee
and
\be
\Box B -  {1 \over \sqrt{6}} \coth
\bigg({ B\over \sqrt{6}} \bigg)
 \barg^{mn} \partial_m B  \partial_n B \
= 0 .
\label{fourd2}
\ee
The five dimensional equations are obtained by substituting the
metric \Ref{Ametric} into the equations \Ref{fiveeinstein}.  The
($mn$) and ($55$) Einstein equations reduce to
\be
\barR_{mn} = - {3 \over 2} 
\left({e^{2 \sigma} \over 1 + A  e^{2 \sigma}}\right)^2
  \partial_m A \partial_n A
\label{five1}
\ee
and
\be
\Box A = {e^{2\sigma}\over 1+A  e^{2\sigma}} \barg^{mn} 
\partial_m A \partial_n A .
\label{five2}
\ee
The ($m5$) equation is satisfied for any value of the radion field.

Equations \Ref{five1} and \Ref{five2} show that the Kaluza-Klein
reduction is {\it not} consistent as it stands.  The fields $\barR$
and $A$ depend only on $x^0,...,x^3$, so eqs.\ \Ref{five1}
and \Ref{five2} cannot be satisfied point by point in $x^5$.
In fact, they must be averaged over the fifth dimension.  This
gives
\bea
\barR_{mn} &=& 2\Lambda \bigg({\keff \over \kappa}\bigg)^2
\int_0^{\pi\rc} dz e^{-2 \sigma} \barR_{mn}   \nn\\[1mm]
&=&- {3 \over 2} {e^{2\pi\rc\Lambda} \over (1 + A)(1 + A
 e^{2 \pi\rc\Lambda})} \partial_m A   \partial_n A \nn\\[2mm]
&=&- {1 \over 4}  \sinh^{-2}\bigg({B\over \sqrt{6}} \bigg)
 \partial_m B \partial_n B
\eea
and
\bea
0&=& 2\Lambda \bigg({\keff \over \kappa}\bigg)^2
\int_0^{\pi\rc} dz e^{-2 \sigma} \left({e^{2\sigma}
\over 1+A  e^{2\sigma}}\right)^2\left(
\Box A - {e^{2\sigma}\over 1+A  e^{2\sigma}} \barg^{mn} 
\partial_m A \partial_n A \right)\nn\\[2mm]
&=& - \sqrt{{3\over2}}\left[
\Box B -  {1 \over \sqrt{6}} \coth
\bigg({ B\over \sqrt{6}} \bigg)
 \barg^{mn} \partial_m B  \partial_n B \right] ,
\eea
in which case they reduce to \Ref{fourd1}
and \Ref{fourd2}, respectively.

This apparent inconsistency has an important physical origin.
It stems from the fact that the higher Kaluza-Klein modes
cannot be set to zero.  Their equations of motion are
such that $\barh_{mn}$ and $\barh_{55}$ acquire tadpoles
proportional to derivatives of the light fields,
\be
\barh_{mn} \sim \partial_m B  \partial_n B , \qquad\qquad
\barh_{55} \sim \barg^{mn} \partial_m B  \partial_n B .
\label{tadpoles}
\ee
These tadpoles alter eqs.\  \Ref{five1} and \Ref{five2} and
restore the consistency of the Kaluza-Klein
reduction.

In general, these tadpoles also change the low energy
effective action.  In this paper, however, we work to {\it second
order} in spacetime derivatives.  To this order, the $\barh_{mn}$
and $\barh_{55}$ tadpoles can be neglected.  The $\barh_{m5}$
field can induce the only dangerous terms.  However, for our
ansatz, the $(m5)$ Einstein equation is always satisfied, so
$\barh_{m5}$ does not acquire a tadpole. Therefore \Ref{SeffR}
is indeed the consistent low energy effective action.

In the rest of this section, we find the Kaluza-Klein reduction
for the graviphoton, $B_M$.  As above, we start with the five
dimensional equation of motion,
\be
\partial_M \left( e g^{MN} g^{PQ} F_{NQ}\right)
 = 0 .
\ee
In the radion background \Ref{Ametric}, this reduces to
\bea
0 &=& \partial_m \left[ \bare  e^{-2 \sigma} (1 + A  e^{2 \sigma})^2
 \barg^{mn}  F_{n5}  \right]  \nn\\[2mm]
0 &=& \partial_5 \left[ \bare  e^{-2 \sigma} (1 + A  e^{2 \sigma})^2
  \barg^{pq} F_{5q}  \right]  +\
\partial_m \left[ \bare (1 + A  e^{2 \sigma})^{-1}
  \barg^{mn}\barg^{pq} F_{nq} \right]  .
\eea

We first find a solution for constant radion.  We take
\be
B_m =  0 ,\qquad\qquad
B_5  =  { \alpha  e^{2 \sigma} \over \kappa (1 + A  e^{2
\sigma})^{2}}
 C  ,
\ee
where
\be
\alpha = {(1 + A)(1 + A  e^{2 \pi\rc\Lambda})\over
(e^{2\pi\rc\Lambda}-1)}
\ee
and $C$ is the gauge-invariant Aharonov-Bohm phase around the fifth
dimension,
\be
C = 2\kappa\Lambda \int_0^{\pi\rc} dz B_5 .
\ee
{}From these expressions it is easy to calculate the field strengths.
We find
\be
F_{m5} = { \alpha  e^{2 \sigma} \over \kappa 
(1 + A  e^{2 \sigma})^{2}} \partial_m C
\label{f1}
\ee
and
\be
F_{mn} = 0 .
\label{f2}
\ee
For constant $A$, eqs.~\Ref{f1} and \Ref{f2} satisfy the
five dimensional equations of motion, provided $C$ is 
a massless scalar field, satisfying $\Box C = 0$.

Let us now fluctuate the radion field.  We assume, for the moment, that
the curvatures \Ref{f1} and \Ref{f2} do not change when $A$ depends
on $x^0,...,x^3$.  If we substitute these expressions into the five
dimensional action,
\be
S  =  - {\Lambda \over 4} 
\int d^5x  e g^{MP} g^{NQ} F_{MN} F_{PQ} ,
\ee
we find
\bea
S &=& - {\Lambda \over 2} 
\int d^5x  \bare  e^{-2 \sigma}(1 + A  e^{2 \sigma})^{2}
 \barg^{mn} F_{m5}  F_{n5}\nn\\
&=&   - {\Lambda \over 2 \kappa^2} 
\int d^5x  \bare  e^{2 \sigma} \left[{
(1 + A)(1 + A  e^{2 \pi\rc\Lambda})
\over(1 + A e^{2 \sigma})(e^{2\pi\rc\Lambda}-1)}\right]^2
 \barg^{mn} \partial_m C  \partial_n C \nn\\[2mm]
&=&    - {1 \over 2 \keff^2}\
\int d^4x  \bare  e^{2 \pi \rc \Lambda}  {(1 + A)
(1 + A  e^{2 \pi\rc\Lambda})\over(e^{2\pi\rc\Lambda}-1)^2}
 \bar{g}^{mn} 
\partial_m C  \partial_n C \nn\\[2mm]
&=&   - {1 \over 8\keff^2} \int d^4x  \bare \
\sinh^{-2}\bigg({B\over \sqrt{6}} \bigg)\
\barg^{mn}  \partial_m C   \partial_n C .
\eea
This is a very suggestive result.  Comparing with \Ref{SeffR}, we
see that the bosonic part of the effective action takes the usual
K\"ahler form,
\bea
S_{\rm eff} &=& \int d^4x  \bare \bigg[ - {1 \over 2 \keff^2}
\barR  - {1 \over 8\keff^2} 
\sinh^{-2}\bigg({ B\over \sqrt{6}} \bigg)  \barg^{mn}\
(\partial_m B  
\partial_n B  + \partial_m C   \partial_n C)
\bigg]\nn\\[2mm]
&=& \int d^4x  \bare \bigg[ - {1 \over 2 \keff^2}
\barR - K_{T\barT}  \barg^{mn}
\partial_m T   \partial_n \barT
\bigg] ,
\label{Seffbosonic}
\eea
where $\keff T = (B + \I C)/\sqrt{2}$, and $\barT$ is its complex
conjugate.  In this expression $K_{T\barT}$ is the K\"ahler
metric
\be
K_{T\barT} = {1\over4}  \sinh^{-2}
\bigg[{ \keff(T+\barT) \over 2\sqrt{3}} \bigg] ,
\ee
derived from the K\"ahler potential\footnote{This K\"ahler
potential was also derived in \cite{LS}.}
\be
K(T,\barT) = -  {3\over\keff^2}  \log
\left(1 - e^{-\keff (T + \barT)/\sqrt{3}}\right) .
\label{kahler}
\ee
Under these assumptions, the Kaluza-Klein reduction gives rise to
a bosonic effective action with K\"ahler potential \Ref{kahler}.

This motivates us to choose an ansatz for $B_M$ that preserves
$F_{m5}$, even for fluctuating $A$.  We take
\bea
B_5 &=& { \alpha  e^{2 \sigma} \over
\kappa (1 + A  e^{2 \sigma})^{2}}
 C \nn\\[2mm]
B_m &=& \int_0^{x^5} d|z| \partial_m  \left[{ \alpha  e^{2 \sigma}
\over \kappa (1 + A  e^{2 \sigma})^{2}}\right]
  C  ,
\label{bs}
\eea
where, as before, $\sigma = \Lambda |x^5|$.  It is a short
exercise to show that \Ref{bs} gives the same $F_{m5}$.  
The ansatz also induces a nonvanishing $F_{mn}$, a
contribution that
can be safely ignored because it is of higher order in
the derivative expansion.

Note that our expression for $B_m$ is odd, as required.  Also
note that it is globally defined, in the sense that the integral
could have run from either $0$ or $\pi\rc$ to $x^5$ (the
difference integrates to zero).  The nonvanishing $B_m$
is a tadpole induced by the higher Kaluza-Klein excitations
of the graviphoton field.

\section{The fermionic effective action}

In the previous section we found that the bosonic part of the
four dimensional effective action is of
K\"ahler form.  In this section we derive the effective
action for the fermions.  We shall see that it is also of
K\"ahler form, as expected for a supersymmetric theory.

Our starting point is the effective theory of the supergravity
multiplet presented in Ref.~\cite{ABN}.  In that paper,
the radion multiplet was set to zero, with $A = C = \Psi_5 = 0$.
The zero mode gravitino was found to be
\be
\psi^+_m = {1\over \sqrt 2} 
\left({\keff\over\kappa}\right) e^{-\sigma/2} 
\psi_m ,\qquad
\psi^-_m = {1\over \sqrt 2} 
\left({\keff\over\kappa}\right) e^{-\sigma/2} 
\sgn(z) \psi_m ,
\label{psizero}
\ee
where the four dimensional gravitino $\psi_m$ depends on the
coordinates $x^0,...,x^3$.

The four dimensional effective action for the supergravity
multiplet was derived by substituting the expressions \Ref{gzero}
and \Ref{psizero} into \Ref{totalS} and integrating over $x^5$.
This gives
\be
S_{\rm eff}  = \int d^4x
 \bare \bigg[  - {1 \over 2 \keff^2} \barR
 +  \epsilon^{mnpq} \bare_n{}^a
 \overline\psi_{m} \bar{\sigma}_{a} D_p \psi_q  \bigg] ,
\label{SeffSUSY}
\ee
plus four-Fermi terms.  Equation \Ref{SeffSUSY} is the $N=1$ 
supergravity action in four dimensions, with an effective four
dimensional squared Planck mass $\keff^{-2}$.

The Randall-Sundrum background preserves one four dimensional
supersymmetry.  For the case at hand, this supersymmetry is
generated by the following Killing spinors,
\be
\eta^+ = {1\over\sqrt2}  e^{- \sigma/2} \eta ,
\qquad\qquad
\eta^- = {1\over\sqrt2}  e^{- \sigma/2} \sgn(z) 
\eta ,
\label{etazeromode}
\ee
where the spinor $\eta$ is a function of $x^0,...,x^3$, but not
$x^5$.  The effective four dimensional supersymmetry transformations
can be found by substituting these zero mode expressions into the
supersymmetry transformations \Ref{susytrans}.  All $x^5$-dependent
terms cancel, leaving
\bea
\delta \bare_m{}^a &=& \I\keff  (\eta \sigma^a\bar
\psi_{m}  +  \bar\eta \bar\sigma^a\psi_{m}) \nn\\[1mm]
\delta \psi_m &=& {2\over\keff} D_m\eta ,
\eea
up to three Fermi terms.  These are the four dimensional $N=1$
supersymmetry transformations  under which the efffective
action is invariant.

We now wish to repeat this exercise in the radion background.
We start by computing the five dimensional equations of motion
for the fermion fields.  We find
\bea
\bar\sigma^n \hat{D}_{5}\psi_{n}^+
 - \bar\sigma^n \hat{D}_{n}\psi_{5}^+
 + {\Lambda\over2} e_{5\hat{5}} \bar\sigma^n\psi^-_n
 +  2\I\Lambda \bar\psi_5^+
 - {\kappa\over\sqrt{6}} g_{55} F^{n5}\bar\psi_n^-\, - &&\nn\\
 - {\I\kappa\over2\sqrt{6}} e_{\hfive 5} 
F^{n5} \bar\sigma_n\psi_5^+
 - \kappa \sqrt{2\over3} 
F_{m5} \bar\sigma^{mn}\bar\psi_n^- &=&  0 \nn\\[2mm]
\bar\sigma^n \hat{D}_{5}\psi_{n}^-
 -  \bar\sigma^n \hat{D}_{n}\psi_{5}^-
 + {\Lambda\over2} e_{5\hat{5}} \bar\sigma^n\psi^+_n
 -  2\I\Lambda \bar\psi_5^-
+{\kappa\over\sqrt{6}} g_{55} F^{n5} \bar\psi_n^+ \, -&&\nn\\
 - {\I\kappa\over2\sqrt{6}} e_{5\hat{5}}
F^{n5} \bar\sigma_n\psi_5^-
 - \kappa\sqrt{2\over3} 
F_{m5} \bar\sigma^{mn}\bar\psi_n^+
 - 2\bar\sigma^n\psi_n^+ [ \delta(z)-\delta(z-\pi\rc) ] &=&
0 \nn\\[2mm]
\sigma^{mn} \hat{D}_m\psi^\pm_n
 \pm {3\I\Lambda\over4} \sigma^n \bar\psi_n^\pm
 - {\I\kappa\over4}  \sqrt{3\over2}  e_{5\hat{5}} 
F^{5n}  \psi_n^\pm &=& 0 .\nn\\
\label{gravitinoeom}
\eea
In this expression, all fields are five dimensional,
$\sigma^m=\sigma^a e_a{}^m$, and we drop terms
that depend on the field strength $F_{mn}$ and the off-diagonal vierbein
elements $e_m{}^\hfive$ and $e_5{}^a$.  The covariant derivatives are
\bea
\hat{D}_{5}\psi_{n}^\pm &=& D_{5}\psi_{n}^\pm
 \mp {\I\over2} \omega_{5a\hfive}\sigma^a\bar\psi^\mp_{n}
 - \I\kappa\Lambda \sqrt{3\over2} B_{5}\psi^\mp_{n} \nn\\
\hat{D}_{n}\psi_{5}^\pm &=& D_{n}\psi_{5}^\pm
 \mp {\I\over2} \omega_{na\hfive}\sigma^a\bar\psi^\mp_{5}
 - \I\kappa\Lambda \sqrt{3\over2} B_{n}\psi^\mp_{5}  \nn\\
\hat{D}_m\psi^\pm_n &=& D_m\psi^\pm_n
 \mp {\I\over2}  \omega_{ma\hat{5}} \sigma^a\bar\psi_n^\mp
 - \I\kappa\Lambda \sqrt{3\over2} B_m \psi_n^\mp .
\eea
All spin connections are fully five dimensional.

In the radion background, we take our ansatz to be
\bea
\psi_n^- &=& \sgn(z)  \psi^+_n \nn\\[3mm]
\psi^+_n &=& {1\over\sqrt{2}} \left({\keff\over \kappa}\right)\left[
 V  W  \psi_n  - {\I\alpha \over \sqrt{6}}
  V^{-5}  W \sigma_n  \bar\chi \right]\nn\\[1mm]
\psi^-_5 &=& \sgn(z) \psi^+_5 + {2 \alpha\over\sqrt{3}} 
\left({\keff\over \kappa}\right)
  e_{5\hat{5}} V^{-5} W^* \chi .
\label{fermiansatz}
\eea
In these expressions,
\be
V = e^{-\sigma/2} (1+A  e^{2\sigma})^{1/4}
\ee
is the fermion warp factor and
\be
W = \exp\left[  \I\kappa \Lambda  \sqrt{3\over2}
\int_0^{x^5} d|z| B_5 \right]
\ee
is a Wilson line along the fifth direction.  Our ansatz
satisfies the fermionic equations of motion for
constant $A$ and $C$.  The Wilson line
reproduces the $x^5$-dependent gauge transformations
of the five dimensional fermions; the zero mode fields are
gauge invariant.
Note that the difference between integrating from $0$ or from
$\pi\rc$ is an $x^5$-independent phase.  This phase can be
absorbed by a K\"ahler transformation in the four dimensional
theory.

Let us first derive the $\chi$ equation of motion.  Subtracting
the first two equations in \Ref{gravitinoeom} and substituting
\Ref{fermiansatz}, we find
\bea
0 &=&\I\alpha^2 \left({e^{2\sigma}\over 1 + A e^{2\sigma}}\right)^2 
\bar\sigma^m D_m\chi + {\I\over2} \partial_m\left[ \alpha^2 
\left({e^{2\sigma}\over 1 + A e^{2\sigma}}\right)^2 \right] 
\bar\sigma^m \chi \, +\nn\\
&& + \kappa \alpha^2 
\sqrt{3\over2} \left( {e^{2\sigma}\over 1 + A e^{2\sigma}}\right)^2
  \left( \Lambda\int_0^{x^5} d|z|  F_{m5} +\
{1\over6} e^{\hfive 5} 
F_{m5}\right) \bar{\sigma}^m 
\chi \, +\nn\\[2mm]
&& + {1\over2} e^{2\sigma} \alpha 
\left( \sqrt{3\over2} \partial_m e_{5\hat{5}}
 - \I\kappa F_{m5} \right)
\bar\sigma^n\sigma^m\bar\psi_n .
\label{eom0.5}
\eea
In these expressions, we take $\sigma^m = \sigma^a \bare_a{}^m$
and we evaluate the spin connections in the radion background,
\bea
\omega_{nab} &=& \bar\omega_{nab} + (e_a{}^m e_{nb} -
e_b{}^m e_{na}) {\partial_m A e^{2\sigma}\over2(1+A e^{2\sigma})}
\nn\\[2mm]
\omega_{na\hat{5}} &=& \Lambda \sgn(z)  e_{na}\nn\\[2mm]
\omega_{5a\hat{5}} &=&
- {\partial_m A e^{2\sigma}\over(1+A  e^{2\sigma})^2} e_a^{~m} ,
\label{fermieom}
\eea
where $\omega_{5ab}=0$ and $\bar\omega_{mab}$ is the four dimensional
spin connection with respect to $\bare_m{}^a$.

Equation \Ref{eom0.5} does not hold point-by-point in $x^5$, so
the ansatz \Ref{fermiansatz} is inconsistent.  As before, the
four dimensional equations of motion can be found by averaging
over the fifth dimension.  This gives
\bea
0&=&\int_0^{\pi\rc}
dz e^{-2\sigma} \Biggl\{
\I\alpha^2 \left({e^{2\sigma}\over 1 + A e^{2\sigma}}\right)^2
 \bar\sigma^m D_m\chi
 + {\I\over2}  \partial_m\left[ \alpha^2 
\left({e^{2\sigma}\over 1 + A e^{2\sigma}}\right)^2 \right] 
\bar\sigma^m \chi \,+\nn\\[2mm]
&& + \kappa \alpha^2 
\sqrt{3\over2} \left( {e^{2\sigma}\over 1 + A e^{2\sigma}}\right)^2
  \left(\int_0^\sigma d\sigma'  F_{m5} + {1\over6} e^{\hfive 5} 
F_{m5}\right) \bar{\sigma}^m 
\chi\, +\nn\\[2mm]
&& + {1\over2} e^{2\sigma} \alpha 
\left( \sqrt{3\over2} \partial_m e_{5\hat{5}}
 - \I\kappa  F_{m5} \right)
\bar\sigma^n\sigma^m\bar\psi_n\Biggr\} ,
\eea
which implies
\bea
0 &=& \I \bar{\sigma}^m D_m\chi
 + {\I\over2} (K^{T\bar{T}} 
\partial_m K_{T\bar{T}}) \bar{\sigma}^m \chi
 + {\keff\over\sqrt{2}} \partial_m \barT 
\bar\sigma^n\sigma^m\bar\psi_n\,+ \nn\\[2mm]
&&+ {1\over 6}  \sqrt{3\over2} 
\left[{1-(2+A) e^{2\pi\rc\Lambda}\over 1-e^{2\pi\rc\Lambda}}\right] 
\partial_m C \bar{\sigma}^m\chi \nn\\
&=& \I \bar{\sigma}^m \tilde{D}_m\chi
 + {\keff\over\sqrt{2}} \partial_m \barT 
\bar\sigma^n\sigma^m\bar\psi_n .
\label{eom1}
\eea
Here
\be
\tilde{D}_m\chi  =  \partial_m\chi  +
 {1\over2} \bar{\omega}_{mab} \sigma^{ab} \chi +\
\Gamma^T_{TT} 
\partial_m T \chi - {\keff^2\over4} 
(K_T\partial_m T - K_{\barT}\partial_m\barT) \chi
\ee
is the covariant derivative, with K\"ahler connection
\be
\Gamma_{TT}^T = - {\keff\over\sqrt{3}}
 \coth\left[{\keff(T+\barT)\over 2\sqrt{3}}\right] .
\ee
Equation \Ref{eom1} is the effective four dimensional $\chi$
equation of motion.

Let us now derive the four dimensional gravitino equation
of motion.  Using the third equation in \Ref{gravitinoeom},
together with the previous results, we find
\bea
0 &=& \epsilon^{mnpq} 
\biggl[ \bar\sigma_n D_p\psi_q  + \I\kappa \sqrt{3\over2} 
\biggl(\Lambda\int_0^{x^5} d|z|  F_{p5}  -  {1\over2} 
e^{\hfive 5} F_{p5} \biggr)  \bar\sigma_n\psi_q\biggr]\ -\nn\\[2mm]
&& - {\alpha\over2} e^{2\sigma} 
\left( \sqrt{3\over2} \partial_n e_{5 \hat{5}}
  +  \I\kappa  F_{n5}\right)
  \sigma^n\bar\sigma^m\bar\chi
\eea
where, as before, $\sigma^m = \sigma^a \bare_a{}^m$.
Averaging over the fifth dimension gives
\bea
0 &=& \int_0^{\pi\rc} dz \Biggl\{ e^{-2\sigma}
 \epsilon^{mnpq} \biggl[ \bar\sigma_n D_p\psi_q
 + \I\kappa \sqrt{3\over2}  \biggl(
\int_0^\sigma d\sigma' F_{p5} - {1\over2} e^{5\hat{5}} F_{p5}
\biggr) \bar\sigma_n\psi_p \biggr] \ -\nn\\[1mm]
&& - {\alpha\over2}
\left(\sqrt{3\over2} \partial_n e_{5 \hat{5}}
  +  \I\kappa F_{n5}\right)
 \bar\sigma^n\sigma^m\bar\chi\Biggr\}  .
\eea
This implies
\bea
0 &=& \epsilon^{mnpq} \bar\sigma_n \biggl[  D_p\psi_q
 - {\I\over2}\sqrt{3\over2} {1+A e^{2\pi\rc\Lambda}
\over e^{2\pi\rc\Lambda}-1} 
\partial_p C \psi_q \biggr]\ - \nn\\[2mm]
&& - {\alpha\over2} \left({\keff\over\kappa}\right)^2 
\partial_n (B+\I C)
 \bar\sigma^n\sigma^m\bar\chi \nn\\[2mm]
&=&\epsilon^{mnpq}  \bar\sigma_n \tilde{D}_p\psi_q
 - {\keff\over\sqrt{2}} 
K_{T\bar{T}} \partial_n T
 \bar\sigma^n\sigma^m\bar\chi ,
\label{eom2}
\eea
where the covariant derivative is
\be
\tilde{D}_m\psi_n  = \partial_m\psi_n  +
 {1\over2} \bar{\omega}_{mab} \sigma^{ab} \psi_n +\
{\keff^2\over4}  (K_T\partial_m T -
K_{\barT}\partial_m\barT) \psi_n  .
\ee
Equation \Ref{eom2} is the equation of motion for the gravitino
$\psi_m$.

The fermionic equations of motion \Ref{eom1} and \Ref{eom2}, together
with their bosonic partners \Ref{fourd1} and \Ref{fourd2}, are
precisely the equations of motion that follow from the following
four dimensional effective action,
\bea
S_{\rm eff} &=& \int d^4x  \bare \bigg[ - {1 \over 2 \keff^2}
\barR + \epsilon^{mnpq} \bar\psi_m\bar\sigma_n
\tilde{D}_p \psi_q\, -\nn\\[2mm]
&&\qquad\qquad - K_{T\barT}  \barg^{mn}  \partial_m T  
\partial_n \barT
 - \I K_{T\barT} \bar\chi\bar\sigma_m \tilde{D}_m\chi\, -\nn\\[3mm]
&&\qquad\qquad -  {\keff\over\sqrt{2}} K_{T\barT} 
\partial_n\barT \chi\sigma^m\bar\sigma^n\psi_m
 - {\keff\over\sqrt{2}} K_{T\barT} 
\partial_n T \bar\chi\bar\sigma^m\sigma^n\bar\psi_m \bigg] .
\label{Sefffull}
\eea
In fact, we have carried out the nontrivial cross-check to show that
\Ref{Sefffull} can be obtained by substituting \Ref{Ametric},
\Ref{bs} and \Ref{fermiansatz} into \Ref{totalS} and integrating
over $x^5$.

\section{Supersymmetry transformations}

In the previous section we found the effective action for the
radion supermultiplet.  We saw that it was of supersymmetric form,
with K\"ahler potential \Ref{kahler}.  In this section we
derive the four dimensional supersymmetry transformations directly
from the five dimensional transformations \Ref{susytrans}.

The derivation of the supersymmetry transformations is more
subtle than the derivation of the effective action.  The action
involves an integral over $x^5$, but the supersymmetry
transformations hold point-by-point in $x^5$.  Therefore, to
derive the supersymmetry transformations, it is helpful to
choose an ansatz in which the radion does not mix with the
graviton, at any point in $x^5$.  Our expression \Ref{Ametric}
has precisely this property.

{}Using our ansatz, it is straightforward to derive the
supersymmetry transformation laws.  We start with the Killing
spinors,
\be
\eta^+ = {1\over\sqrt2}  V W \eta ,
\qquad\qquad
\eta^- = {1\over\sqrt2}  V W \sgn(z) 
\eta
\label{etafullzeromode}
\ee
which parameterize the one unbroken supersymmetry in the radion
background. We substitute \Ref{Ametric}, \Ref{bs}, \Ref{fermiansatz}
and \Ref{etafullzeromode} into \Ref{susytrans} and cancel all
$x^5$ dependence.  In this way we derive the transformation
laws
\bea
\delta \bare_m{}^a &=& \I\keff  (\eta \sigma^a\bar
\psi_{m}  +  \bar\eta \bar\sigma^a\psi_{m}) \nn\\[1mm]
\delta \psi_m &=& {2\over\keff} \tilde{D}_m\eta \nn\\[1mm]
\delta T &=& \sqrt{2}  (\eta\chi  +  \bar\eta\bar\chi)\nn\\[1mm]
\delta \chi &=& \I \sqrt{2} \sigma^m \bar\eta
 \partial_m T ,
\eea
under which \Ref{Sefffull} is invariant.  In these transformations,
the covariant derivative of $\eta$ is given by
\be
\tilde{D}_m\eta  = \partial_m\eta  +
 {1\over2} \bar{\omega}_{mab} \sigma^{ab} \eta +\
{\keff^2\over4}  (K_T\partial_m T - K_{\barT}\partial_m\barT) \eta
 .
\ee

Let us demonstrate how this works for the case of the radion multiplet.
(The transformation laws for the supergravity multiplet are found
using the same technique.)  We first compute the variation of
$T$,
\bea
\delta T &=& {1\over \sqrt{2} \keff} (\delta B  +  \I \delta C) \nn\\
&=& {\Lambda\over\keff} \int_0^{\pi\rc} dz (\sqrt{3}  \delta
e_{5\hfive}
 +  \I\sqrt{2} \kappa \delta B_5) \nn\\
&=& 2\sqrt{3}  \Lambda \left({\kappa\over\keff}\right) 
\int_0^{\pi\rc} dz (\eta^+\psi_5^-  -  \eta^-\psi_5^+) \plushc\nn\\
&=& 2\sqrt{2}  \alpha   \Lambda 
\int_0^{\pi\rc} dz  V^{-4}  e_{5\hfive} 
(\eta\chi  +  \bar\eta\bar\chi)\nn\\
&=& \sqrt{2}  (\eta\chi  +  \bar\eta\bar\chi) ,
\eea
where we drop all three Fermi and higher derivative terms.
In a similar way, we compute the variation of $\chi$,
\bea
\delta \chi &=& {\sqrt{3}\over 2 \alpha}\left({\kappa\over\keff}\right)
 e^{\hfive 5} V^5 W (\delta\psi^-_5  - 
\sgn(z) \delta\psi^+_5)\nn\\
&=& - {\I\over\keff} \sigma^m \bar\eta \left(\sqrt{3\over2}
 {1\over\alpha} \partial_m A - \I \partial_m C \right)\nn\\
&=& \I \sqrt{2} \sigma^m \bar\eta \partial_m T .
\eea
All $x^5$ dependence cancels, leaving the
supersymmetry transformations of the low energy
four dimensional effective theory.

\section{Conclusions}

In this paper we derived the effective action and supersymmetry
transformations for the radion supermultiplet in the supersymmetric
Randall-Sundrum scenario with two opposite tension branes.  We
found the action to be of the standard supersymmetric form, with
K\"ahler potential
\be
K(T,\barT) = -  {3\over\keff^2}  \log
\left(1 - e^{-\keff (T + \barT)/\sqrt{3}}\right) .
\label{kp}
\ee
The supersymmetric action is the starting point for studies of radion
stabilization in supersymmetric theories \cite{LS}.

The K\"ahler potential \Ref{kp} has the correct limit as the warp
factor vanishes.  Indeed, rescaling $T \rightarrow \Lambda T$,
$\chi \rightarrow \Lambda \chi$ and taking $\Lambda \rightarrow 0$,
it is not hard to see that the effective action reduces to the usual
no-scale form, with K\"ahler potential
\be
K(T,\barT) \rightarrow -  {3\over\keff^2}  \log(T + \barT) .
\ee

It is perhaps worth noting that a simple argument suggests the
above form for the K\"ahler potential.  The argument proceeds as
follows:

\begin{enumerate}

\item
Start with the four dimensional supergravity action, with four
dimensional Planck constant $\keff^{_2}$, where $\keff^{-2}
= \kappa^{-2} (1 - e^{-2 \pi \rc\Lambda})$.  Then transform to
the supergravity frame, in which the Einstein term takes the
following form,
\be
S_{\rm sugra}  = -  {1 \over 2 \keff^2}\int d^4x
 \bare e^{- \keff^2  K(T,\barT)/3} \barR .
\label{sugra}
\ee
The K\"ahler potential $K(T,\barT)$ is an unknown real function
of the complex radion field $T$.

\item
Note that for constant radion, the K\"ahler potential is itself
a constant, so the action \Ref{sugra} describes Einstein gravity
with an effective Planck constant $\keff^{-2} \exp(-\keff^2 K/3)$.
The K\"ahler potential must vanish when $T$ is set to its expectation
value, $\langle T + \barT \rangle = 2\sqrt{3}\pi\rc\Lambda/\keff$.

\item
Now shift the radion expectation value, $\langle T + \barT \rangle
\rightarrow 2\sqrt{3}\pi\rc'\Lambda/\keff$.  According to \Ref{sugra},
this changes the effective four dimensional Planck constant from
$\keff^{-2}$ to $\keff^{\prime  -2} = \keff^{-2} \exp(-\keff^2
K(T',\barT')/3).$
Consistency then requires $\bar{\kappa}^{\prime  -2} = \kappa^{-2}
(1 - e^{-2 \pi \rc'\Lambda})$.

\item
Combine these results to find
\be
K(T,\barT) = -  {3\over\keff^2} \log\left[
{1 - e^{-\keff(T+\barT)/\sqrt{3}} \over
1 - e^{-2 \pi \rc\Lambda}}\right] .
\ee
Up to a constant, this is precisely the K\"ahler potential \Ref{kp}.
\end{enumerate}

\section*{Acknowledgments}

During the course of this work, we have benefited from discussions
with many people, including Richard Altendorfer, Kiwoon Choi, Adam
Falk, Ruth Gregory, Emanuel Katz, Horatio Nastase, Erich Poppitz, Yael
Shadmi and Yuri Shirman.  We would also like to thank Markus Luty
and Raman Sundrum for extensive discussions, and point out that
they independently deduced the K\"ahler potential \Ref{kp} in
Ref.~\cite{LS}.

We acknowledge the hospitality of the Aspen Center for Physics,
and financial support from the National Science Foundation,
grants NSF-PHY-9970781 and NSF-PHY-0070928, and the Department
of Energy, contract DE-FG03-84ER40168.

\end{document}